# Surfactant-induced-retardation in lateral migration of droplets in a microfluidic confinement

Somnath Santra[a], Sayan Das[a], Sankha Shuvra Das[a], and Suman Chakraborty[a,*]

[a]*Department of Mechanical Engineering, Indian Institute of Technology Kharagpur, Kharagpur – 721302, India*

**ABSTRACT**

In the present study, the role of surfactants on the cross-stream migration of droplets is investigated both experimentally and theoretically. For experimental analysis, sunflower oil is used as the carrier phase and DI water as the dispersed phase which is intermixed with Triton X-100 that acts as the surfactant. A T-junction is used in the microchannel for the purpose of droplet generation. Presence of an imposed pressure driven flow induces droplet deformation and disturbs the equilibrium that results in subsequent surfactant-redistribution along the interface. This further induces a gradient in the surface tension, thus generating a Marangoni stress that significantly alters the droplet dynamics. On subsequent experimental investigation, it is found that presence of surfactants reduces the cross-stream migration velocity of the droplet. Further, it is shown that the effect of surfactants in reducing the cross-stream migration is significantly enhanced for a larger droplet as compared to a smaller one within the same time span, provided the channel height is kept constant. In addition, a larger surfactant concentration is found to induce a greater retardation in cross-stream migration of the droplet, the effect of which is reduced when the initial transverse position of the droplet is shifted closer to the channel centerline. The present analysis can be applied to various droplet based microfluidic as well as medical diagnostic devices where manipulation of droplet trajectory is a major issue. For the theoretical counterpart, an asymptotic approach is adopted in the presence of bulk-insoluble surfactants and under the assumption of negligible fluid inertia. A good match between our theoretical prediction and the experimental results is obtained.

\* E-mail address for correspondence: suman@mech.iitkgp.ernet.in

# 1. INTRODUCTION

The study of the dynamics of droplets in a microchannel has significant importance due to its wide variety of applications in the domain of microfluidics, for instance, different droplet based microfluidic devices that include processes such as drug delivery, cell encapsulation, analytic detection, single cell analysis and positioning of erythrocytes in blood flow through arteries [1–7]. Surface active agents or surfactants, which may be present naturally or may be artificially added, are quite common in microfluidic devices and are, in general, utilized for the purpose of manipulating different polymeric properties as well as stabilizing various emulsions [8–10]. A fundamental understanding as well as proper control of the position of the dispersed phase in a emulsion provides a wide scope in the domain of flow field fractionation [11,12] and flow cytometry [13].

Migration of a droplet in the presence of a pressure driven flow has been studied both theoretically and experimentally since a long time [14–16]. Amongst them, the different experimental techniques used can be classified into two categories [17]: the first one involves the use of additional laminar flows to locate a laminar stream containing particles and the second one uses hydrodynamic lift forces to drive particles in a cross-stream direction (which is normal to the direction of local flow velocity). The experimental methodology used for the present study falls under the second category. These hydrodynamic lift forces can be produced as a result of variety of mechanisms [18–23]. Out of all the lift forces, the inertial lift force deserve special mention. Even though inertial lift forces in microfluidic devices is significantly less, Di Carlo et al. [6,24] used the same as a tool to control the position of particles in a microchannel. In a recent study, Stan et al. [14] studied experimentally the cross-stream migration of droplets in a pressure driven microflow. In their study, they took into consideration the effect of buoyancy and also the lift force induced due to droplet deformation and channel confinement. A number of previous studies can be found in the literature that deals with lift forces due to droplet deformation [18,19] or due to hydrodynamic interaction between droplets or bubbles with channel walls [25–28]. Presence of surfactants, on the other hand, adds further nonlinearity to the physical system and hence significantly affects the droplet dynamics. Variation in surfactant concentration results in a subsequent alteration of interfacial tension that gives rise to a Marangoni convection and hence a Marangoni stress, which happens to be the source of this nonlinearity [29–31]. This Marangoni stress in conjunction with other lift forces, mentioned above, brings in new dynamics related to cross-stream migration of a droplet, which is the prime motivation of our present study.

A good number of theoretical investigation on lateral migration of droplet has also been previously carried out. Amongst them, Haber and Hetsroni [32] developed a three-dimensional theoretical model on the migration of an eccentrically placed droplet in a arbitrary flow field, with the aid of Lambs general solution. Chan and Leal [18] analytically studied the migration of a deformable droplet in an imposed Poiseuille flow and showed that a droplet always migrates towards the flow centerline when the viscosity of the dispersed phase is significantly low as compared to the carrier phase. Kim and Subhramaniam [30] developed a two-dimensional axisymmetric theory that investigated the role of a temperature gradient on the axial migration of a surfactant-laden droplet for the limiting case of both low and high surface Péclet number ($Pe_s$), which signifies the relative importance of surface convection-dominated surfactant transport with respect to surface diffusion-dominated surfactant transport. They also developed a numerical model for arbitrary $Pe_s$. In a recent study, Pak et al. [31] analytically investigated the effect of surfactants on the cross-stream migration of a non-deformable droplet in the limiting regime of low $Pe_s$. As the present study takes into account both the effect of droplet deformation and surfactant redistribution on its cross-stream migration, a subsequent temptation of application of linear superposition of the results of Pak et. al [31] and Chan & Leal [18] will be misleading due to unknown shape of the droplet. This challenge was tackled earlier in the work done by Das et al. [33] and a similar approach is adopted for the present study as well, for the case of an imposed plane Poiseuille flow.

In the current literature, although a significant amount of research has been performed related to surfactant-induced-Marangoni stress, none the less a thorough experimental analysis regarding the effect of surfactants on the lateral migration of droplets in a microchannel is still lacking. Motivated by the same, we, in the present study, focus on the effect of surfactant distribution on the lateral migration of a droplet from an experimental perspective. Rigorous experiments are performed with sunflower oil as the carrier phase and water as the dispersed phase along with Triton X-100 as surfactant. Our experiments are primarily directed towards showcasing the effect of channel confinement ratio (which is the ratio of the droplet radius to the channel height), surfactant concentration and droplet initial position on its cross-stream migration. It is observed that a larger confinement ratio enhances the retarding effect of surfactants as compared to a system with smaller confinement ratio. It is also found that even though a higher surfactant concentration increases the retardation in cross-stream migration of the droplet, the influence of the same significantly reduces if the initial transverse position of the droplet is shifted more close to the channel centerline. In addition, a three-dimensional theoretical model that looks into the effect of surfactants on cross-stream migration of a deformable is also provided to support our experimental observations. It is seen that there is a good match between our theoretical predictions and the outcomes from experiments performed for a microchannel with low confinement ratio and in the presence of a dilute surfactant concentration.

## 2. THEORETICAL APPROACH

For the theoretical model, a physical system is chosen that comprises of a neutrally buoyant surfactant-laden droplet of radius *a*, suspended in an unbounded plane Poiseuille flow field. The droplet is initially placed at an off-center position with respect to the channel centerline. For the purpose of an theoretical analysis, the droplet radius is assumed to be much smaller in comparison to the channel height, $\bar{H}$, that is the confinement ratio, $a/\bar{H} \ll 1$ which effectively signifies an unbounded flow field. Surfactants, insoluble in either of the phases, are present at the interface of the droplet. Presence of an imposed Poiseuille flow results in droplet deformation accompanied by surfactant re-distribution that causes a variation in surface tension along the interface and hence generates Marangoni stress that plays an important role in the cross-stream droplet migration. The theoretical model derived can predict the temporal variation of the transverse position of a deformable droplet in the presence of surfactants.

Some of the basic assumptions, relevant to our experiments, are negligible fluid inertia, small droplet deformation, surface diffusion dominated surfactant transport and a dilute surfactant concentration along the interface. The first assumption clearly indicates a low flow Reynolds number, $Re = \rho \bar{V}_c a / \mu_e \ll 1$, where $\rho$ and $\mu_e$ is the carrier phase density (918 kg/m$^3$) and viscosity (0.4914 Pa-s at 25$^\text{o}$ C) and $\bar{V}_c$ is the centerline velocity of the imposed pressure driven flow. The next two assumptions indicate a low value of capillary number ($Ca = \mu_e \bar{V}_c / \sigma_{eq}$) and surface Péclet number ($Pe_s = \bar{V}_c a / D_s$) where $\sigma_c$ is the surface tension at the fluid-fluid interface and $D_s$ is the surface diffusivity of the surfactant used, which for the present scenario of Triton X-100 has magnitude of $D_s \sim 10^{-6}$ m$^2$/s. Based on the material properties, all of these assumptions are found relevant to the experiments performed in this study.

Under the premise of the above assumptions, the flow field is governed by the Stokes and the continuity equations which can be expressed as

$$\left.\begin{array}{l} -\bar{\nabla}\bar{p}_i + \mu_i \bar{\nabla}^2 \bar{\mathbf{u}}_i = \mathbf{0}, \ \bar{\nabla} \cdot \bar{\mathbf{u}}_i = 0, \\ -\bar{\nabla}\bar{p}_e + \mu_e \bar{\nabla}^2 \bar{\mathbf{u}}_e = \mathbf{0}, \ \bar{\nabla} \cdot \bar{\mathbf{u}}_e = 0, \end{array}\right\} \tag{1}$$

where $\bar{p}, \bar{\mathbf{u}}$ represent the pressure and velocity field and $\mu$ denotes the dynamic viscosity. Subscripts '$i$' and '$e$' refer to dispersed phase and the carrier phase respectively. The surfactant concentration is governed by the surfactant transport equation at the droplet interface and is given by

$$\bar{\nabla}_s \cdot \left( \bar{\mathbf{u}}_s \bar{\Gamma} \right) = D_s \bar{\nabla}_s^2 \bar{\Gamma}, \tag{2}$$

where $\bar{\nabla}_s$ is the surface gradient operator and $\bar{\Gamma}$ is the local surfactant concentration. The flow field governing equations (eq (1)) are subjected to kinematic and stress balance conditions at the interface as well as far-field conditions, the details of which are provided in the supplementary material. As seen from eq (1) and (2), the governing equations for flow field and the surfactant transport are coupled through the surface convection term in the convection-diffusion equation for surfactant concentration. In addition, presence of shape deformation brings in further nonlinearity into the system as the shape of the droplet is not known as a priori. As a result, an exact analytical solution for an arbitrary value of $Pe_s$ is impossible. We thus utilize the asymptotic approach to solve for the flow field and surfactant concentration. A regular perturbation method is applied under the limiting case of surface-diffusion driven surfactant transport or $Pe_s \ll 1$. Such a limiting case is close to a real scenario owing to low characteristic velocities [$O(10^{-4})$ m/s] of flow in a microchannel. The methodology adopted for the present study is similar to that used by Das et al. [33], where the bulk flow was taken to be a circular Poiseuille flow. Hence the same is not repeated in this article. However, the necessary details and expressions are provided in the supplementary material for reference. The final expression for cross-stream migration velocity is provided in the results and discussion section.

## 3. EXPERIMENTAL SETUP AND METHODOLOGY

We now discuss on the experimental setup as well as the methodology adopted in the present study. All the experiments are performed in a controlled environment at 25 $^{o}$C to prevent any influence of thermal fluctuations on fluid flow. In our experiments, commercially available sunflower oil is used as the carrier phase while water along with Triton X-100 is used as dispersed phase. Triton X-100 acts as the surfactant that alters the interfacial tension in presence of imposed pressure driven flow which ultimately leads to the generation of Marangoni stress.

### A. MICROCHANNEL FABRICATION

A conventional methodology (which consists of photo-lithography technique followed by soft lithography technique) is used for PDMS microchannel fabrication [34]. Initially, the substrate (essentially a glass slide/silicon wafer) is cleaned with piranha solution ($H_2O_2 : H_2SO_4 = 1:1$) for ~1 hour followed by DI water rinsing. To remove moisture from the substrate, it is then dried by purging by pure $N_2$ gas followed by heating at 95°C for about 1 hour. Then, the substrate is coated with SU8-2150 negative photoresist (Micro Chem Corp., USA) at 3000 rpm for about 25 sec. Further the photoresist coated substrate is soft-baked at 65°C and 95°C for 7 mins and 40 mins respectively. Thereafter, the substrate is then exposed under UV light (Hybralign 200 Mask aligner, OAI) through a required chrome-coated quartz mask for about 25 sec. However, to further harden the exposed area, the substrate is then post-baked at 65°C and 95°C for 5 min and 15 mins respectively. Finally, the substrate is developed using SU8 developer for about 17 mins to get the exposed pattern (or master pattern) on the top of the substrate.

To prepare the microchannel, PDMS (polydi-methylsiloxane), a elastomeric resin is prepared by mixing elastomeric base to cross-linker (Sylgard-184, Dow corning, USA) in 10:1 w/w ratio. Then, the degassed PDMS is poured over the master pattern and is cured in the hot plate at 95 °C overnight. After complete curing, the PDMS channel is peeled-off and bonded with a clean glass surface using $O_2$-plasma treatment.

### B. EXPERIMENTAL SETUP

We now discuss on the experimental setup. A detailed schematic of the same is provided in figure 1. The setup includes an Olympus IX71 inverted fluorescent microscope which is fitted with a high speed camera (Phantom V641) to capture the necessary images. The PDMS based microchannel, which is plasma-bonded on the glass slide comprises of three inlet and one outlet ports, all of which are fitted with Teflon tubes of identical diameters through connectors.

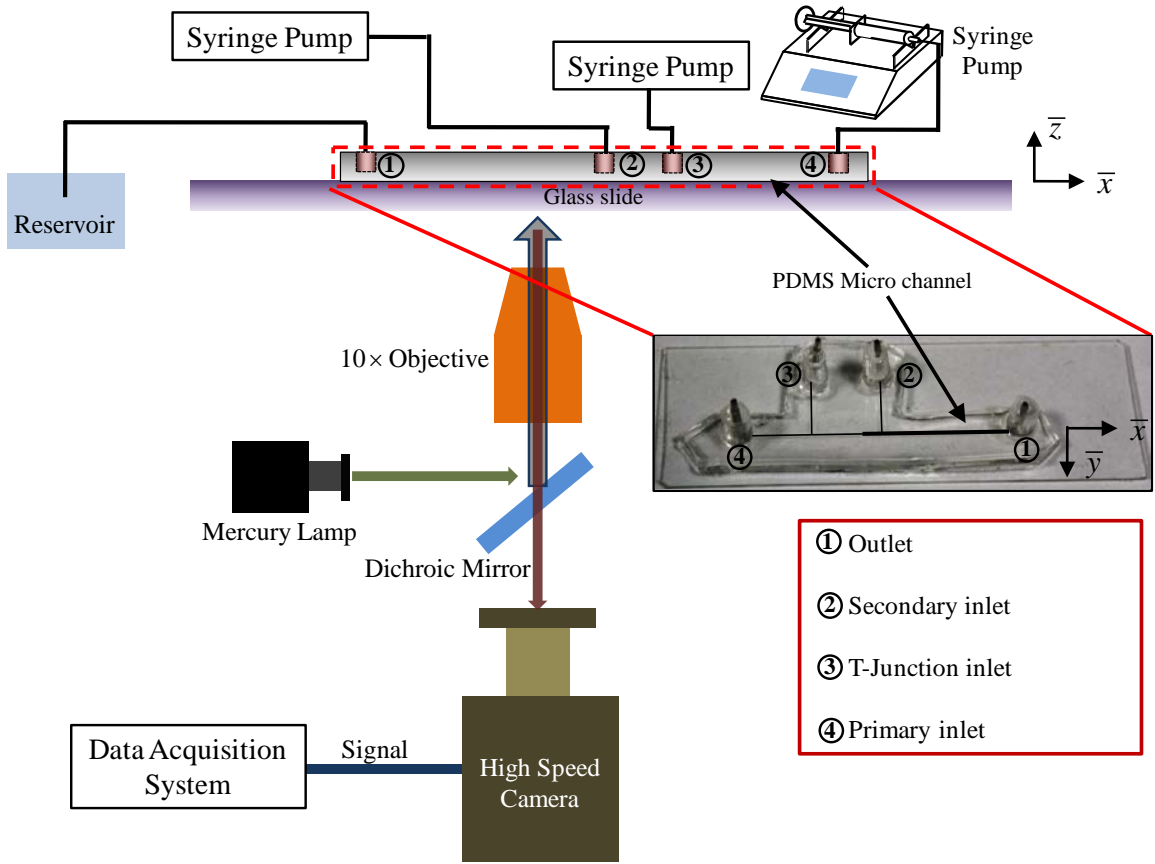

Figure 1. A schematic of the experimental setup used for the present analysis. The cartesian coordinate $(\bar{x}, \bar{y}, \bar{z})$ attached to the bottom plate of the mircochannel is also shown in the figure above.

The three inlet Teflon tubes consists of a primary inlet which causes the inflow of sunflower oil, a T-junction inlet that forces entry of water intermixed with Triton X-100 and finally the secondary inlet which also results in the entry of sunflower oil. Each of these three Teflon tubes are connected to separate syringe pumps (Harvard Apparatus PHD 2000: 0-100 ml min$^{-1}$) that forces the respective fluid at a desired flow rate whereas the outlet is connected to a reservoir. The microchannel is placed on the observation platform of the inverted microscope such that the portion of the channel with enlarged width just past the secondary inlet lies above the 10x objective.

## C. METHODOLOGY

Unlike some of previous studies performed in this domain where a flow-focusing device was used for the purpose of droplet generation [14,35–37], we have made use of a T-junction for the same purpose. The use of a T-junction significantly reduces the complexity in channel fabrication without any sacrifice in the level of accuracy. A priming test is performed with a test liquid (Milli-Q ultrapure water; Millipore India Pvt. Ltd) prior to performing real time experiments to check for any leakage in the channel. Subsequent inflow of sunflower oil from the primary inlet and water intermixed with Triton X-100 from the auxiliary inlet (T-junction) is initiated at flow rates of 120 μl/hr and 20 μl/hr respectively for a 200 μm channel.

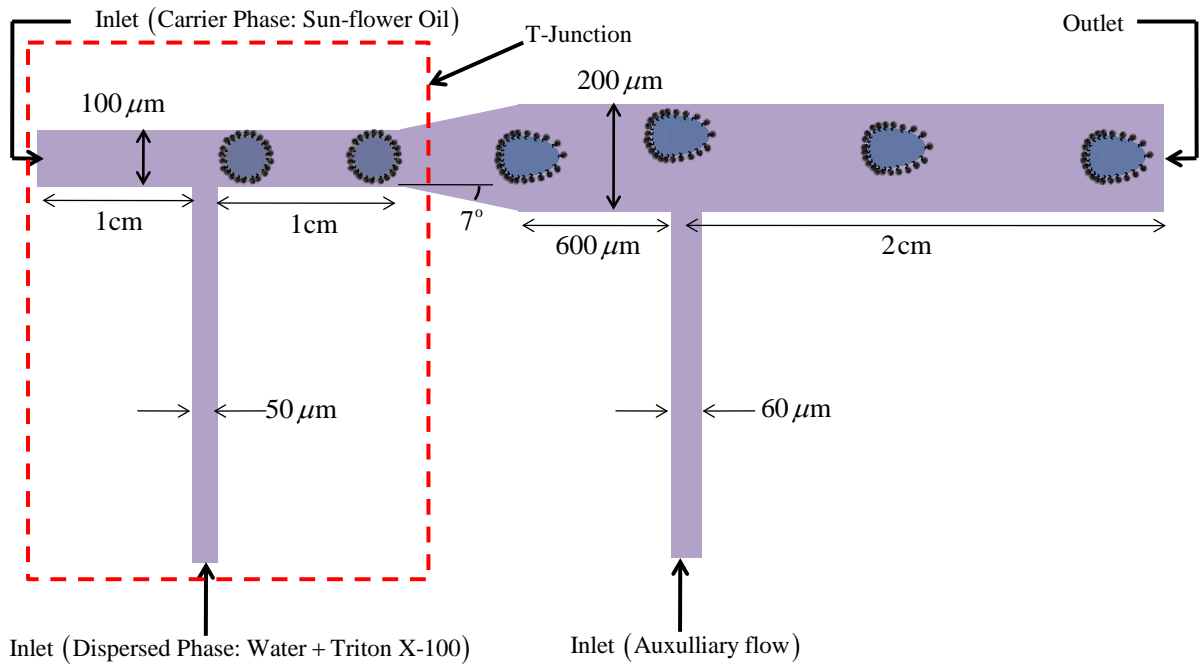

Figure 2. A schematic of the migration of a surfactant laden droplet in the microchannel. The red box region, highlighted above, represents the T-junction responsible for droplet generation.

As soon as droplet generation starts, the flow of sunflower oil through the secondary inlet is also initiated at a rate of 15 μl/hr in order to offset the position of the droplet with respect to the channel centerline. The above experiment is performed with 800 ppm of Triton X-100 intermixed with DI water. The experiment is repeated again for different concentrations of Triton X-100 in the 200 micron channel. Subsequent experiments are also performed for different droplet radius in a 300 micron channel to study the effect of confinement ratio or bounding wall on the lateral migration of the droplet. In addition, by altering the secondary flow rate of sunflower oil, the offset position of the droplet was varied in microchannels of height 500 μm to investigate the influence of this initial offset position of the droplet on its trajectory. The entire trajectory of the droplet past the secondary inlet is recorded by means of a high speed camera.

For accurate visualization of the droplet migration in the flow field, images (image resolution: $600\times800$ pixels $\times 12$ bits) are captured at intervals of $\Delta t = 10^{-3}$ secs at the rate of 1000 frames/s and at the same exposure. Post-processing of the images captured are done with the help of an in-house image processing code in MATLAB. A schematic of the flow field in the microchannel is presented in figure 2. All the experiments in our setup are performed in a plane, orthogonal to the direction of gravitational force. As a result, there is no effect of buoyancy on the droplet dynamics irrespective of the difference in density of either of the phases.

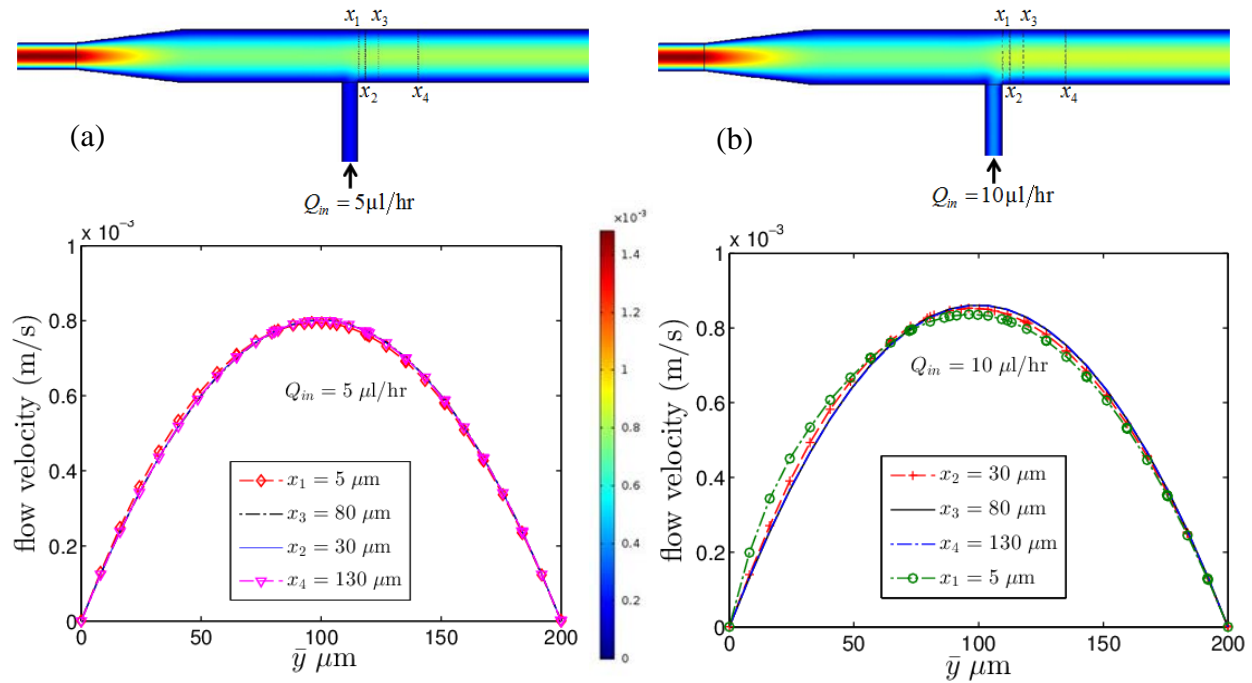

Figure 3. Plot of velocity profiles at different sections, $x_1$, $x_2$, $x_3$ and $x_4$, (measured from the secondary inlet) obtained by performing a numerical simulation of the flow field. The color bar indicates the magnitude of fluid flow velocity. Figure (a) shows the velocity profiles when the secondary influx is $Q_{in}$ = 5 μl/hr whereas in Figure (b) the corresponding influx is increased to $Q_{in}$ = 10 μl/hr.

It should be noted that our area of interest lies in the region past the secondary inlet and hence the transverse position of the droplet as result of this influx is taken as the initial position of the droplet trajectory. To be more precise, the initial position of the droplet is its location downstream the secondary inlet where the streamlines reach a steady state indicating a fully developed flow and all other disturbances alienate. A more clear overview can be provided if we look into the velocity profiles at different sections ($x_1$, $x_2$, $x_3$, $x_4$) in this region as shown in the figure 3, obtained after performing a numerical simulation of the flow field. The distance of each sections, shown in legends of the plots, are measured from the secondary inlet. It can be seen from figure 3(a) that the velocity profiles at sections $x_2$, $x_3$, $x_4$ overlap each other which suggests that the flow field becomes fully developed from $x_2 = 30$ µm when the secondary inflow rate is $Q_{in} = 5$ µl/hr. On the other hand, figure 3(b) shows that the velocity profiles overlap only after $x_3 = 80$ µm, that is the flow is fully developed after $x_3 = 80$ µm when $Q_{in} = 10$ µl/hr. It can thus be inferred that higher the secondary inflow rate, higher is the transverse initial position, which at the same time gets shifted further downstream. This axial distance of the initial position varies with change in channel height as well. Hence for each of the experiments, a corresponding numerical simulation is performed to acquire the axial distance of the droplet's initial position, from which point all observations on droplet dynamics are recorded.

## 4. RESULTS AND DISCUSSIONS

In this section, we first demonstrate the role played by surfactants on the temporal variation of the transverse position of the droplet. The droplet, when suspended in an imposed flow field undergoes migration and at the same time due to interfacial fluid flow, the surfactants gets redistributed. This leads to a non-uniform distribution of surfactants along the droplet surface resulting in an asymmetry in its concentration on either sides of both the axial as well transverse planes. This non-uniform distribution of surfactants generates a gradient in surface tension along the interface and hence gives rise to a Marangoni stress. Towards showcasing the effect of surfactants on the cross-stream migration of a droplet, we first plot the temporal variation in the experimentally measured transverse position of the droplet centroid in a 500 µm microchannel and compare the same with the theoretical prediction for the case of an unbounded flow field (refer to figure 4(b)). In figure 4(b), experimental data points for the lateral position of the droplet centroid are shown for the case of a clean droplet and a surfactant-laden droplet (100 ppm). The radius of the droplet for the present case is set at 50 µm by appropriately varying the flow rate of the dispersed phase ($Q_d$) from the auxiliary inlet for a constant inflow rate of the carrier phase ($Q_c$). The theoretical prediction for the transverse migration of the droplet as a function of time can be derived from the expression of the cross-stream migration velocity of the droplet which is given by

$$U_y = Ca \left[ \underbrace{(c_1 y_d + c_2)}_{\substack{\text{sole effect due to} \\ \text{shape deformation}}} + \underbrace{\frac{\sum_{i=1}^{4} \left( \sum_{j=1}^{4} c_{5-i,5-j} k^{5-j} \right) \beta^{5-i}}{c_3}}_{\substack{\text{combined effect of shape deformation} \\ \text{and surfactant redistribution}}} \right], \quad (3)$$

where the constant coefficients $c_1$, $c_2$, $c_3$ and $c_{p,q}$ ($p,q \in [1,4]$) are provided in the section 2 of the supplementary material. Here $\beta$ is the elasticity parameter that signifies the sensitivity in surface tension to a change in surfactant concentration along the droplet surface and $k = Pe_s/Ca$ is the property parameter. The detailed expression of either of these parameters are provided in the supplementary material. The above expression clearly indicates the role of shape deformation as well as the imposed flow induced surfactant-redistribution on the cross-stream migration velocity of the droplet. The expression also indicates that cross-stream migration of the droplet takes place even without the presence of surfactants. The strategy used in order to obtain the expression for cross-stream migration velocity [eq (3)] is similar to that used by Das et al [cite]. The left hand side of the above equation can be replaced by $U_y = dy_d/dt$, which provides us with an ordinary differential equation in $y_d$. The solution of eq (3) for $y_d$, thus obtained, is plotted in figure 4(b) with and without the presence of surfactants. The time, $t$, along the $x$- axis is normalized with respect to the characteristic time given by $a/\overline{V}_c$. As seen from the same figure, in the absence as well as in the presence of surfactants, our asymptotic theory succeeds in predicting the experimental results to a great extent. The values of the different parameters used for the above plot are provided in the caption of the figure 4. The value of $\overline{\Gamma}_{eq}$ corresponding to a surfactant concentration of 100 ppm is calculated and is found to be $5 \times 10^{-6}$ Mol/m$^2$. The corresponding magnitude of the elasticity number for an imposed flow rate of approximately $10^{-10}$ m$^3$/s and a droplet radius of 50 μm is found to be $\beta = 0.5$. For the analytical plot in figure 4, the value of capillary number is taken as 0.1, which is justified based on different material properties. Among other parameters, the viscosity ratio is calculated to be $\lambda = 0.018$ and the property parameter, $k$ (= 1.35) is used as a fitting parameter to obtain a good match between the theoretical and the experimental results.

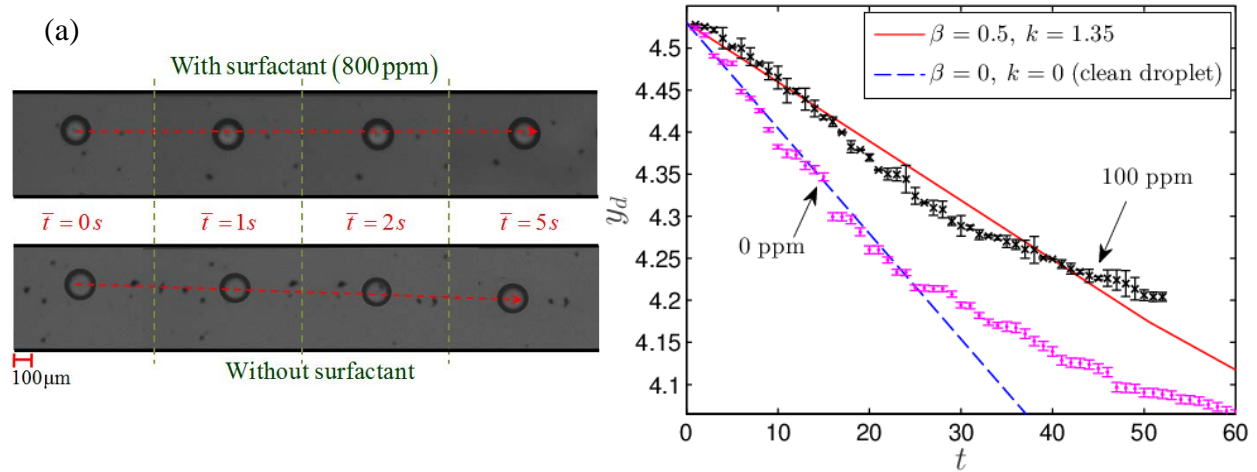

Figure 4. Figure (a) shows the trajectory of a surfactant-laden as well as a clean droplet in a 500 μm microchannel, recorded by a high speed camera. Figure (b) shows the variation of the transverse position of the droplet as function time elapsed, in the presence (100 ppm) as well in the absence of surfactants. The solid and dashed lines indicate the theoretical prediction with ($\beta = 0.5$, $k = 1.35$) and without ($\beta = k = 0$) the presence of surfactants. The other parameters used for the theoretical plot are $\lambda = 0.018$ and $Ca = 0.1$.

It is evident from the experimental data points in figure 4(b) that the presence of surfactants retards the cross-stream migration of the droplet. The droplet traverses a much smaller distance in the cross-stream direction in the presence of surfactants for the same time elapsed. In other words, the cross-stream migration velocity of the droplet reduces in the presence of surfactants. Figure 4(a) shows the snapshots of the droplet at different time frames as recorded with the help of a high speed camera. Comparing the trajectories of both the surfactant-free as well as surfactant-laden droplet (in figure 4(a)), we infer that cross-stream migration velocity for the former is higher with respect to the later.

A proper physical explanation of the role of surfactants on the lateral migration of a droplet can be provided if we look into its distribution along the interface. The shape deformation and hence the presence of surface divergence vector makes it complicated to show the variation of surfactant concentration along the droplet surface. However the surface divergence vector can be evaluated on the droplet surface when the surfactant concentration is projected on an undeformed spherical surface from an deformed one with the use of the relationship $\tilde{\Gamma} = \Gamma\left(r_s^2/\mathbf{n}\cdot\mathbf{r}\right)$ [38]. This is presented in figure 5(a), which shows a contour plot for the distribution of surfactants projected along the undeformed droplet surface, as obtained from our theoretical prediction for the case of an unbounded flow. In our experiments as well as in our theoretical model, the droplet is positioned at an off-center location with respect to the flow centerline. For the experiments, a secondary inflow is used for the same purpose.

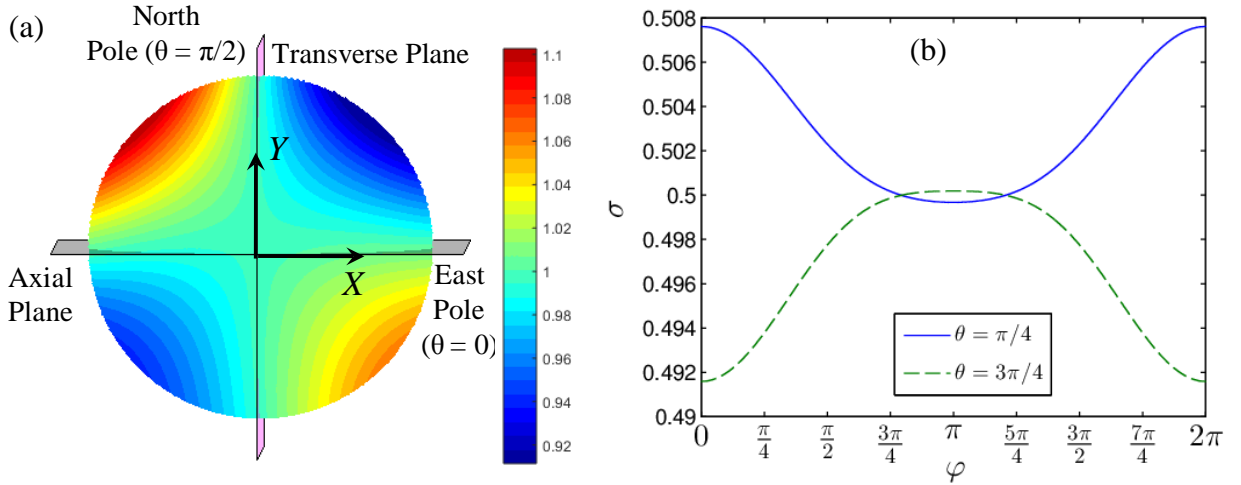

Figure 5. (a) Contour plot showing the distribution of surfactants along the droplet surface. (b) The variation of surface tension along two different axial planes ($\theta = \pi/4$, $3\pi/4$) is shown as a function of the azimuthal angle, $\varphi$. The different parameters involved for the above plots are $H = 4$, $\lambda = 0.018$, $y_d = 2.8$, $Ca = 0.1$, $\beta = 0.5$ and $k = 1$.

Considering the case when the droplet is located at an off-center position above the centerline, it can be stated that the lower hemisphere of the droplet has a higher interfacial fluid flow velocity as compared to that of the above. In addition, due to axial migration of the droplet in the direction of the imposed flow, fluid flow along the surface occurs from the front stagnation point ($\theta = 0$) to the rear stagnation point ($\theta = \pi$) of the droplet. Taking into account either of these factors the highest concentration of surfactant is expected near the north-west region of the droplet surface while the minimum surfactant concentration can be predicted to be near the north-east region. This is evident from figure 5(a) as well. The important thing to be noted from the same figure is the asymmetry in surfactant distribution along the droplet surface, both about the axial as well as the transverse plane. It should be kept in mind that this asymmetry in surfactant concentration is also a result of the imposed flow induced shape deformation of the droplet. This asymmetric surfactant distribution gives rise to a gradient in surface tension. It is this asymmetry in surfactant concentration and hence the gradient in surface tension across the axial plane that affects the cross-stream migration of the droplet. To obtain a better insight, we have also shown the variation of the surface tension ($\sigma = 1- \beta\Gamma$) along two axial planes ($\theta = \pi/4$, $3\pi/4$) on either side of the equator ($\theta = 0$), corresponding to the parameter values provided in the caption of figure 5. Figure 5(b) clearly indicates the gradient in surface tension ($|\sigma_{max} - \sigma_{min}|$) across the axial plane of the droplet which results in the generation of Marangoni stress. It should also be stated that the presence of droplet deformation distorts the fluid flow around it, which generates a hydrodynamic force that tends to drive the droplet towards the channel centerline. The asymmetry in surfactant distribution is further enhanced as a result of shape deformation. The Marangoni stress, thus generated acts against the hydrodynamic force due to the imposed flow induced shape deformation. This is also evident from the contour plot in figure 5(a), where the lowest surface tension can be seen to be present in the region of highest surfactant concentration (north-west zone), which implies that the traction due to the Marangoni stress acts against the imposed hydrodynamic force.

It should be noted that in a previous study by Chan & Leal [18] theoretically investigated the cross-stream migration of a deformable droplet in a pressure driven flow, where as in a recent work by Pak et al. [31] did a similar analysis on a non-deformable droplet in the presence of surfactants. Although either of them showed that the droplet always migrated towards the channel centerline for low viscous droplets, non-the-less a mere linear superposition of their results to obtain our current theoretical prediction would be erroneous. The sole reason for the same is presence of shape deformation which renders the system of governing equations and relevant boundary conditions non-linear, since the shape of the droplet is not known as a priori. We next discuss the effect of some of the pertinent parameters, frequently encountered in microfluidic devices, on the cross-stream migration of the droplet. These parameters include the confinement ratio of the channel, the surfactant concentration in the dispersed phase as well as the droplet initial position. For each of these cases experimental results are shown and proper reasoning is provided regarding the behavior of droplet.

A.  EFFECT OF CHANNEL CONFINEMENT

In this section, we primarily focus on the effect of bounding walls on the surfactant-induced retardation in the cross-stream migration of the droplet. To vary the confinement ratio, $a/\bar{H}$, which is the ratio of the droplet radius to the channel height, we opt to alter the size of the droplet. This is possible by varying the inflow of the dispersed phase through the auxiliary channel ($Q_d$) for a constant inflow rate of the continuous phase ($Q_c$) in the T-junction. A larger inflow rate of the dispersed phase result in a larger volume of the same to pass through the auxiliary channel before the continuous phase cuts the auxiliary inflow at the T-junction. Hence a larger droplet size is attained by increasing the inflow ratio, $Q_d/Q_c$. We, thus, analyze the effect of channel wall on lateral migration of the droplet for three different values of $a/\bar{H}$ (0.295, 0.325, 0.36) corresponding to the T-junction inflow ratios ($Q_d/Q_c$) of 0.153, 0.23 and 0.35 respectively.

Figure 6(a) shows the microscopic images of droplets of varying sizes in a microchannel of height 300 μm. This in turn corresponds to different confinement ratios for the same channel height, $a/\bar{H} = 0.36, 0.325, 0.295$. The variation of the transverse position of a clean droplet as a function of time elapsed for different values of $a/\bar{H}$ is shown in figure 6(b). The time used in this plot is normalized with respect to the characteristic time scale given by $\bar{H}/\bar{V_c}$, since the channel height, $\bar{H}$, is kept constant. It can be observed from the plot that decrease in the confinement ratio causes the droplet to traverse a smaller distance in the cross-stream direction for the same time elapsed. That is, for a constant channel height, a larger droplet possesses a higher cross-stream migration velocity. Similar observations can also be made from figure 6(c) where a surfactant-laden droplet of larger size is seen to traverse a greater distance in the same time interval as compared to a droplet of smaller size. The experimental observation, as shown in figure 6(b) for a clean droplet, are in direct agreement with the numerical and experimental results of Mortazavi & Tryggvason [39] and Stan et al. [14], where they showed that it is the larger droplet that reaches its steady state position (or the channel centerline) in the shortest span of time. An interesting observation can also be made on comparison of figures 6(b) and 6(c). It is seen that effect of surfactants in reducing the cross-stream migration velocity of a larger droplet or a droplet in a system with a higher confinement ratio, is more significant as compared to a droplet of smaller size. In other words, for a constant channel height and for the same amount of surfactants added to the dispersed phase (1200 ppm), the decrease in the distance traversed in the transverse direction is significantly more for a larger droplet as compared to a smaller droplet for the time elapsed. If we look into figure 6(b), the distance ($\bar{y}_d/\bar{H}$) traversed by a larger droplet (say $a/\bar{H} = 0.36$) in the transverse direction in the absence of any surfactant is given by 0.05, whereas that traversed by a smaller droplet (say $a/\bar{H} = 0.295$) is 0.027 for the same time elapsed ($t^* = 5$). This agrees with our former observation. Now from figure 6(c), it can be seen that due to presence of surfactants, the transverse distance traversed by the larger and the smaller droplet ($a/\bar{H} = 0.36, 0.295$) reduces to 0.028 and 0.017, respectively, for the same time elapsed ($t^* = 5$). Hence for the larger droplet, presence of surfactants reduces the transverse distance traversed by 0.022, whereas for the smaller droplet the decrease in the distance traversed in the cross-stream direction due to the presence of same amount of surfactants is 0.01 in the same time span ($t^* = 5$).

A proper physical explanation is now provided on the effect of confinement as discussed above. It can be noted that the hydrodynamic force due to shape deformation, which tends to drive the droplet towards the channel centerline, is greater for a droplet of larger size as compared to a smaller one. This explains the plot shown in figure 6(b), where a surfactant-free droplet, larger in size, migrates in the cross-stream direction at a higher rate as compared to a smaller droplet. On the other hand, a larger droplet has a greater difference in interfacial fluid flow velocities between the upper and lower hemispheres as compared to a smaller droplet in the same microchannel. This results in a greater asymmetry in surfactant distribution for a larger droplet and hence a higher Marangoni stress is generated that opposes the hydrodynamic force due to the imposed flow induced shape deformation.

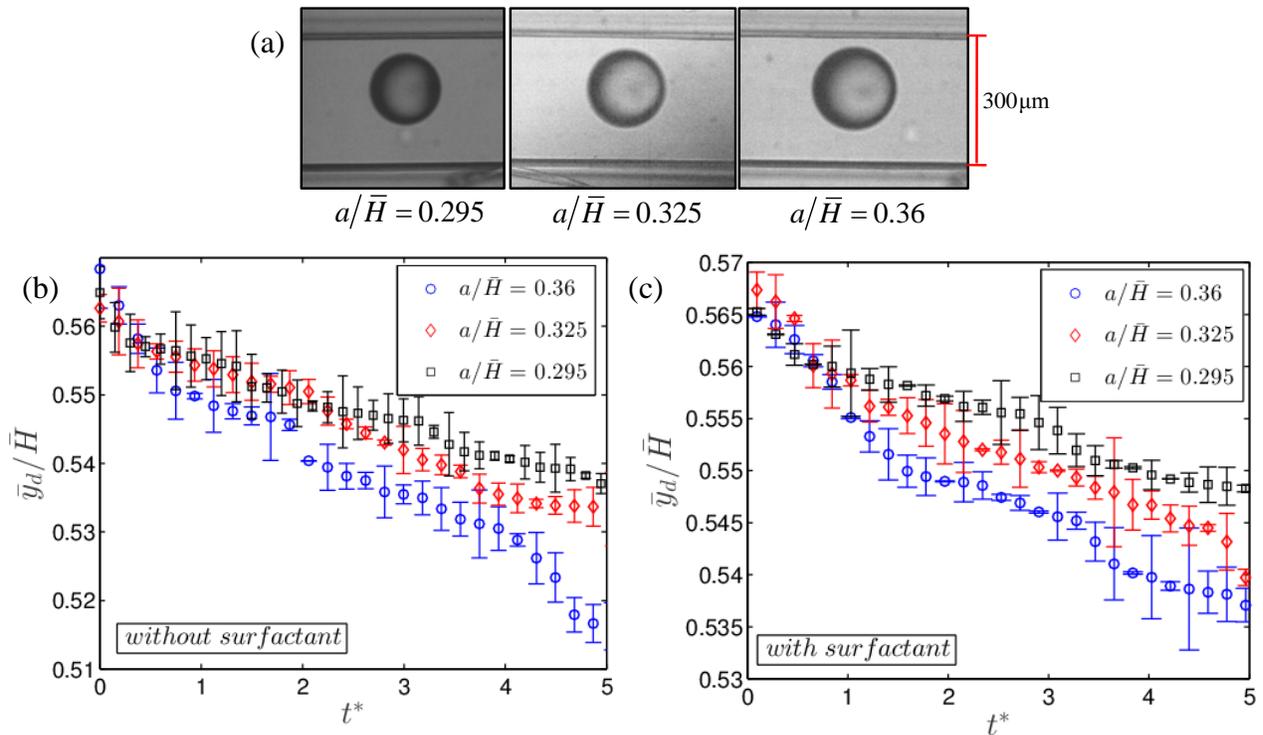

Figure 6. (a) Experimentally recorded images of three different droplets of different size, migrating in a microchannel of height 300 μm. (b) Temporal variation of normalized transverse position of a surfactant-free droplet for different confinement ratios. (c) Variation of transverse position of a surfactant-laden droplet (1200 ppm) as a function of time for different values of confinement ratios (0.295, 0.325, 0.36).

Thus a fall in the net hydrodynamic force, acting on droplets of different sizes, is expected. In other words, the cross-stream migration of a clean droplet is affected more significantly in comparison to a surfactant-laden droplet for the same variation in its size. This can be observed from comparison of figures 6(b) and 6(c) and also has been discussed earlier. In the presence or in absence of surfactants, the larger droplet (or a higher channel confinement ratio) always possesses the highest cross-stream migration velocity. However, as a result of reduction in droplet size (for a constant channel height) by the same amount, the decrease in the cross-stream migration velocity is higher for a surfactant-free droplet as compared to a surfactant-laden droplet.

**B. EFFECT OF SURFACTANT CONCENTRATION**

We next look into the effect of surfactant concentration on the cross-stream migration of the droplet for a constant confinement ratio. Towards this, we perform our experiments corresponding to three different amounts of surfactants (100 ppm, 800 ppm and 1200 ppm) dissolved into the dispersed phase in a microchannel of height 200 μm. Figure 7(a) shows the time variation of transverse position of the droplet while figure 7(b) shows the droplet trajectory. It can be observed from the former that increase in the amount of surfactant concentration reduces the cross-stream migration velocity of the droplet, which will thus take a longer time to traverse the same distance in the cross-stream direction. On the other hand, figure 7(b) shows that the droplet with the highest concentration of surfactant traverses the least in the transverse direction for the same axial distance covered.

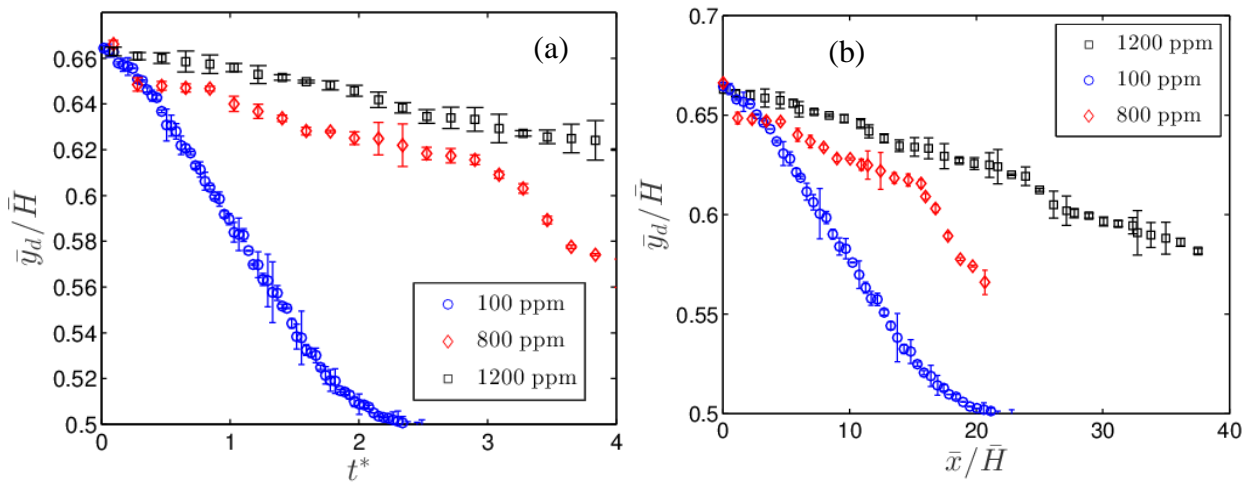

Figure 7. (a) Experimentally observed variation of normalized lateral position of the droplet as a function of normalized time for different values of surfactant concentration (100 ppm, 800 ppm, 1200 ppm). (b) Corresponding droplet trajectory along the length of the microchannel of height 200 μm.

A physical explanation regarding the above nature of droplet dynamics can be put forward if we look into the role played by the surfactants. For the same imposed flow, a droplet with a higher concentration of surfactants will have a greater asymmetry or a higher gradient in surfactant concentration ($|\Gamma_{max} - \Gamma_{min}|$) on either sides of the axial plane. This will result in an increase in the surface tension gradient which in turn will lead to the generation of Marangoni stress of larger magnitude. Since the Marangoni stress generated opposes the hydrodynamic force due to the shape deformation, a droplet with a higher surfactant concentration will have a lower cross-stream migration velocity as compared to a droplet with lower surfactant concentration. In other words, a droplet with lower surfactant concentration will reach the channel centerline faster in comparison to a droplet with a higher surfactant concentration which also implies that the same will traverse a less axial distance. This is what can be observed from figures 7(a) and 7(b).

## C. EFFECT OF DROPLET INITIAL POSITION

We next focus on the effect of droplet initial transverse position on the cross-stream migration of the droplet. How the surfactant-induced retardation of the droplet is affected by altering the droplet initial position, is also another aspect of this discussion. An accurate way to determine the transverse initial position of the droplet has been previously discussed. In order to vary the initial position, we alter the inflow rate of sunflower oil through the secondary inlet. A higher influx results in an initial position further away from the channel centerline. For the present scenario, we choose three inflow rates of 80 μl/hr, 100 μl/hr and 130 μl/hr to achieve the desired initial transverse positions of 275 μm, 315 μm and 365 μm, respectively. A microchannel of height 500 μm is chosen for the experiments to be performed for this analysis.

Figure 8(a) shows the images of three initial transverse position of the droplets, $\bar{y}_{d,0}/\bar{H}(=0.55, 0.63, 0.73)$, where $\bar{y}_{d,0} = \bar{y}_d(t=0)$. Figure 8(b), on the other hand, shows effect of droplet initial position or the secondary influx on the surfactant induced retardation of the droplet cross-stream migration. Interestingly, it can be observed that the closer the droplet is to the channel centerline, the lower is the effect of variation in surfactant concentration on its cross-stream migration. However, if the surfactant concentration of the droplet is fixed, then an increase in its initial transverse position merely increases the time required by it to reach the channel centerline with a subsequent decrease in the cross-stream migration velocity. This can be observed from figure 8(c).

A proper reasoning regarding the above observations is now provided. For a droplet positioned at a point further away from the channel centerline $\left[\bar{y}_d(t=0)/\bar{H}=0.73\right]$, the difference in interfacial fluid flow velocity between its upper and lower hemispheres is larger as compared to a droplet situated near to the centerline $\left[\bar{y}_d(t=0)/\bar{H}=0.63\right]$. This results in a larger asymmetry in surfactant distribution along the droplet surface on either sides of the axial plane and hence a higher surface tension gradient for the former. A larger Marangoni stress is thus generated that opposes the hydrodynamic force due to droplet deformation. Thus increase in the total amount of surfactant along the droplet surface would result in an enhanced Marangoni stress for a droplet situated far away from the channel centerline. This supports our observation from figure 8(b) that for the same increase in surfactant concentration, reduction in the cross-stream migration velocity is larger for a droplet near to the wall. For a fixed surfactant concentration, on the other hand, a higher Marangoni stress is generated for a droplet positioned furthest from the channel centerline which results in a lower cross-stream migration velocity. This explains the fact, as seen from figure 8(c), that a droplet far from the channel centerline would require a larger time to traverse the same distance in the transverse direction as compared to a surfactant-free droplet.

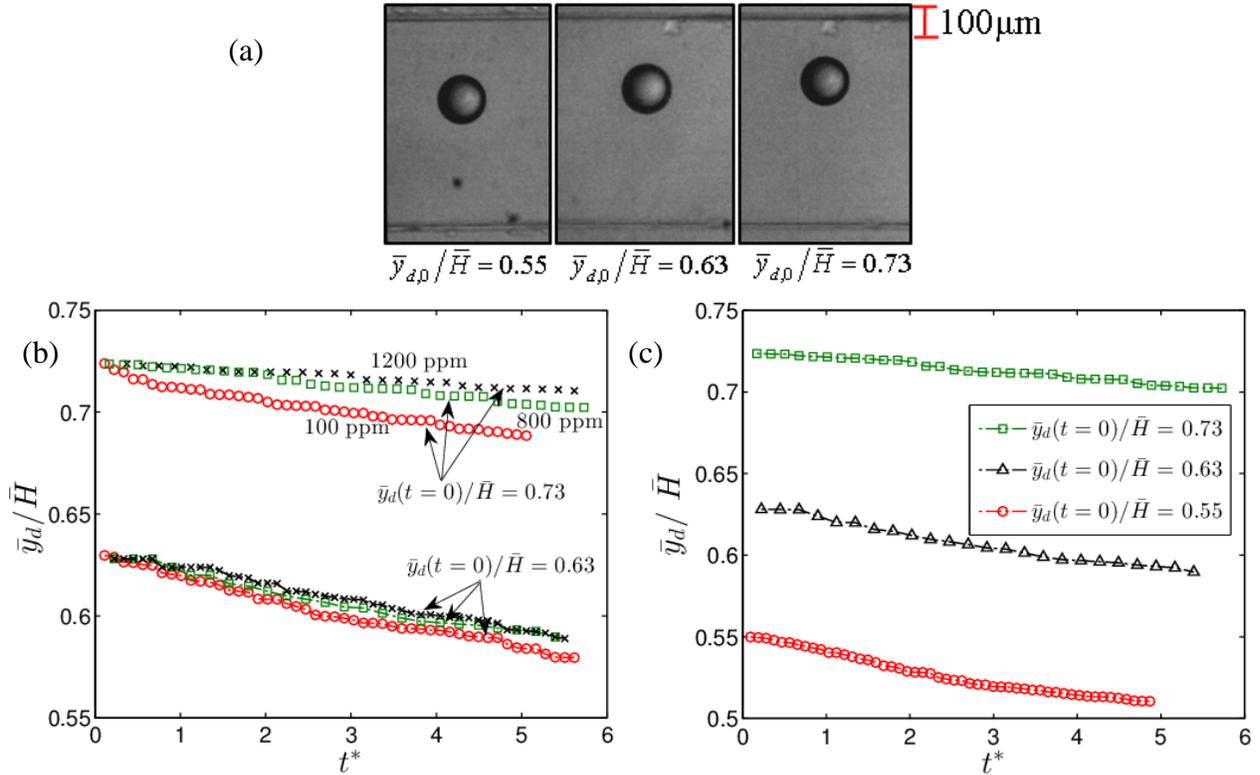

Figure 8. (a) Images of droplet corresponding to three different secondary inflow rates that results in different transverse initial position of the droplet. (b) Variation of normalized transverse position of the droplet as a function normalized time elapsed for two different initial positions $\left[\bar{y}_d(t=0)/\bar{H} = 0.73, 0.63\right]$ and for different surfactant concentrations (100 ppm, 800 ppm, 1200 ppm) for each of these cases. (c) Temporal variation of the lateral position of the droplet corresponding to different initial positions (0.55, 0.63, 0.73) and a total surfactant concentration of 800 ppm.

## 5. CONCLUSION

In the present study, we experimentally show that presence of surfactants in the dispersed phase, suspended in a pressure driven flow, can bring about retardation in its cross-stream migration. Surfactants can be naturally present in a suspension (for example contaminants) or may be artificially added to appropriately modulate the droplet dynamics. We also experimentally investigate the influence of some important parameters commonly encountered in various microfluidic applications, such as the confinement ratio, total concentration of surfactants as well as the initial transverse position of the droplet, on its migration characteristics. In addition, we develop a three-dimensional asymptotic theory under the limiting case of diffusion-dominated-surfactant transport that takes into account the effect of shape deformation as well. Due to the coupled and non-linear nature of the governing equations for flow field and surfactant transport equation, a linear supposition of results doesn't serve the purpose. Instead a more strategic asymptotic approach has been adopted. Overall, our theoretical model is found to predict our experimentally obtained results with a decent accuracy. Some of the noteworthy outcomes of our study are presented below

(i) Irrespective of any change in parameters involved, the time required for a surfactant-laden droplet to reach the channel centerline is always larger as compared to a clean droplet. A good match between our theoretical prediction and the experimental data is found for both the case of a surfactant-free and a surfactant-laden droplet.

(ii) For a fixed channel height and for the same time elapsed, the cross-stream migration of a larger droplet is reduced to a greater extent as compared to a smaller droplet for the same amount of surfactants present.

(iii) A larger amount of surfactant in the dispersed phase is found to significantly reduce the cross-stream migration velocity.

(iv) The effect of surfactant concentration on the cross-stream migration of the droplet is found to be enhanced when the initial position is shifted further away from the channel centerline. However, for a constant amount of surfactant present in the dispersed phase, the droplet initially located near to the channel centerline has a higher cross-stream migration velocity.

**ACKNOWLEDGEMENT**

SS and SD are grateful to Dr. Shubhadeep Mandal for suggesting this problem and to Dr. Aditya Bandopadhyay for his guidance in developing the in-house image processing code in MATLAB. SS and SD are also thankful to Miss. Udita U. Ghosh for her thorough insight on various experimental intricacies.

**Supplementary material**

Please refer to the supplementary material for details of the governing equations and relevant boundary conditions used for the asymptotic analysis along with the expressions of the important results.